\documentclass[12pt]{article}

\usepackage{amsmath}

\begin{document}

\title{Charged analogue of Finch-Skea stars}

\author{S. Hansraj and S. D. Maharaj \\
Astrophysics and Cosmology Research Unit,\\
 School of Mathematical
Sciences, University of KwaZulu-Natal,\\ Durban 4041, South Africa}

\maketitle

\begin{abstract}
We present solutions to the Einstein-Maxwell system of equations in
spherically symmetric gravitational fields for static interior
spacetimes with a specified form of the electric field intensity.
The condition of pressure isotropy yields three category of
solutions. The first category is expressible in terms of elementary
functions and does not have an uncharged limit. The second category
is given in terms of Bessel functions of half-integer order. These
charged solutions satisfy a barotropic equation of state and contain
Finch-Skea uncharged stars. The  third category is obtained in terms
of modified Bessel functions of half-integer order and does not have
an uncharged limit.  The physical features of the charged analogue
of the Finch-Skea stars are studied in detail. In particular the
condition of causality is satisfied and the speed of sound does not
exceed the speed of light. The physical analysis indicates that this
analogue is a realistic model for static charged relativistic
perfect fluid spheres.
\end{abstract}

\section{Introduction}

In the past many classes of exact solutions of the Einstein field
equations have been found for static spherically symmetric
gravitational fields with isotropic matter. A comprehensive list of
Einstein solutions is provided by Delgaty and Lake \cite{Lake}.
These solutions may be used to model a neutral relativistic star as
they are matchable to the Schwarzschild exterior at the boundary. In
comparison fewer exact solutions of the Einstein-Maxwell field
equations are known for static spherically symmetric gravitational
fields with isotropic matter with nonzero electric fields. A
recent review of  Einstein-Maxwell solutions is given by Ivanov
\cite{Ivanov}. These solutions may be utilized to model a charged
relativistic star as they match to the Reissner-Nordstrom exterior.

In modelling a charged relativistic sphere it is desirable to ensure
that two general features are contained in the model. Firstly the
model should be physically reasonable: the gravitational,
electromagnetic and matter variables are continuous and well behaved
in the interior, the interior metric matches smoothly to the
exterior spacetime and causality is not violated. Secondly we should
regain an uncharged solution (which should also satisfy the relevant
physical conditions) of the the Einstein equations when the
electrical field vanishes; a neutral star should be regainable as a
stable equilibrium end state. The anisotropic charged model of
Sharma et al \cite{Sailo} is an example that satisfies the stated
criteria. An approach in satisfying the two criteria is to construct
the model such that the limiting uncharged solution is a known exact
solution. This is not easy to achieve in practice as the number of
known exact uncharged solutions satisfying all conditions of
physical acceptability are limited as established by Delgaty and
Lake \cite{Lake}.

Our objective in the paper is to generate a new solution of the
Einstein-Maxwell system that is physically acceptable and
necessarily contains a neutral stellar model. The neutral stellar
model is the Finch and Skea star \cite{Finch} which satisfies all
the requirement of physical acceptability.
 In \S 2 we express the Einstein equations for neutral matter
 and the Einstein-Maxwell system for charged matter as
equivalent sets of differential equations utilising a transformation
due to Durgapal and Bannerji \cite{Durgapal}. We choose particular
forms for one of the gravitational potentials and the the electric
field intensity in \S 3. The condition of pressure isotropy becomes
 a second order linear equation in the remaining gravitational
potential. We integrate the condition of pressure isotropy and
consequently produce three classes of exact solutions to the
Einstein-Maxwell field equations which can be written explicitly in
terms of elementary functions as shown in \S 3.1, \S 3.2 and \S 3.3.
We regain the uncharged exact solution found previously from the
charged analogue of the Finch and Skea model in \S 3.2; we also
demonstrate that the charged solution satisfies a barotropic
equation of state. In \S 4 we comprehensively study the physical
features of the charged Finch-Skea model. Graphs are generated for
particular parameter values for the gravitational, electromagnetic
and matter variables. We consider in particular the gravitational
behaviour at the centre, and also produce a specific  value for the
the charge to radius ratio. We believe that this detailed analysis
of the matter variables represents a realistic model of a compact
charged object. Some brief concluding remarks are made in \S 5.

\section{Einstein-Maxwell equations}
We assume that the interior of a spherically symmetric relativistic
star is described by the line element
    \begin{equation}
    ds^2 = -e^{2\nu(r)}dt^2 + e^{2\lambda(r)} dr^2
    + r^{2} (d\theta^2 +\sin^{2}\theta \,d\phi^{2}) \label{spherical}
    \end{equation}
in Schwarzschild coordinates $(t,r, \theta,\phi)$ where the
functions $\nu (r)$ and $\lambda (r)$ are gravitational potentials.
For neutral perfect fluids the Einstein field equations can be
expressed as follows
\begin{subequations} \label{perfect}
    \begin{eqnarray}
\frac{1}{r^2}  [r(1-e^{-2\lambda})]'& =& \rho \label{perfect-a}
\\   -\frac{1}{r^2}\ (1-e^{-2\lambda})+\frac{2\nu'}{r}\
e^{-2\lambda}& =& p \label{perfect-b}  \\   e^{-2\lambda}
\left(\nu''+{\nu}'^{2} +\frac{\nu'}{r}  -\nu' \lambda'-
\frac{\lambda'}{r}\right) & = & p \label{perfect-c}
\end{eqnarray}
    \end{subequations}
for the geometry described by (\ref{spherical}). We measure  the
energy density $\rho$ and pressure $p$ relative to the comoving
fluid 4-velocity $u^a = e^{-\nu} \delta^{a}_{0}$ and primes denote
differentiation with respect to the radial coordinate $r$. We
utilize units where the coupling constant $\frac{8 \pi G}{c^4} = 1$
and the speed of light $c=1$.
 An  equivalent form of the field
equations is obtained if we introduce the transformation
\begin{equation}  A^2 y^2 (x) = e^{2 \nu(r)}, \,\,\, Z(x) =
e^{-2 \lambda(r)}, \,\,\, x = C r^2 \label{transf}
 \end{equation}
 where the quantities $A$ and $C$ are
 arbitrary constants. Under
the transformation (\ref{transf}) the system (\ref{perfect}) has the
form
    \begin{subequations} \label{Durg}
    \begin{eqnarray}
    \frac{1-Z}{x}-2\dot{Z} &=&\frac{\rho}{C}            \label{Durg-a}\\
    4Z\frac{\dot{y}}{y} + \frac{Z-1}{x} &=& \frac{p}{C} \label{Durg-b}\\
    4Zx^2 \ddot{y} + 2\dot{Z}x^{2} \dot{y} + (\dot{Z}x-Z+1)y&=&0    \label{Durg-c}
    \end{eqnarray}
    \end{subequations}
where the dots denotes differentiation with respect to the variable
$x$.

A generalisation of the system (\ref{Durg}) is  the coupled
Einstein-Maxwell field equations  given by
\begin{subequations} \label{Durgc}
\begin{eqnarray}
\frac{1-Z}{x} - 2\dot{Z} & = & \frac{\rho}{C} +
 \frac{E^{2}}{2C} \label{Durgc-a}\\
 4Z\frac{\dot{y}}{y} + \frac{Z-1}{x} & = & \frac{p}{C} -
 \frac{E^{2}}{2C}\label{cubic17-b} \label{Durgc-b}\\
 4Zx^{2}\ddot{y} + 2 \dot{Z}x^{2} \dot{y} + \left(\dot{Z}x -
Z + 1 - \frac{E^{2}x}{C}\right)y & = & 0 \label{Durgc-c}\\
 \frac{\sigma^{2}}{C} & = & \frac{4Z}{x} \left(x \dot{E} + E
\right)^{2} \label{Durgc-d}
\end{eqnarray}
\end{subequations}
where $E$ is the electric field intensity and $\sigma$ is the proper
charge density. When the electric field $E=0$ then the
Einstein-Maxwell equations (\ref{Durgc}) reduce to the Einstein
equations (\ref{Durg}) for neutral matter. The system of equations
(\ref{Durgc}) governs the behaviour
 of the gravitational field for a charged perfect fluid.
The particular representation of the Einstein-Maxwell system as
given in (\ref{Durgc}) may be easier to integrate in certain
situations as demonstrated by Thirukkanesh and Maharaj
\cite{Thirukkanesh}. An interior stellar solution of (\ref{Durgc}),
for the line element (\ref{spherical}), should match to the exterior
gravitational field described by the Reissner-Nordstrom line
element. In terms of Schwarzschild  coordinates the
Reissner-Nordstrom solution has the form
\begin{equation}
ds^2=-\left( 1-\frac{2M}{r} + \frac{Q^2}{r^2}\right)dt^2 +
 \left( 1-\frac{2M}{r} + \frac{Q^2}{r^2}\right)^{-1} dr^2 +
r^2(d\theta^2 + \sin^2 \theta d\phi^2) \label{exterior}
\end{equation}
where $M$ and $Q$ are associated with the mass and
charge of the sphere respectively.

\section{Exact Solutions}

It is evident that the system of equations (\ref{Durgc}) is
underdetermined. Consequently we need to specify two of the
variables in advance so that a solution can be obtained. A possible
approach is to specify an equation of state relating $\rho$ to $p$
and additionally to specify one of the gravitational potentials.
This leads to differential equations that are highly nonlinear and
difficult to integrate. Another option, which is pursued here, is to
postulate a form for the gravitational potential $Z$ and to choose
a form for the electrostatic field $E$. Our choice for the function
$Z(x)$ is contained in the form used by Thirukkanesh and Maharaj
\cite{Thirukkanesh}  and Maharaj and Mkhwanazi \cite{MaharajM} which
ensures that we regain as a special case models of fluid spheres
with uncharged matter distributions analysed previously.

The metric function $Z(x)$ is chosen to be of the form
\begin{equation}
 Z = \left( 1 + x \right)^n \label{Z}
 \end{equation}
 where $n
\neq 0$ is a real number. (If $n=0$ then (\ref{Durgc-a}) implies
that $\rho=-E^2/2$ which is negative.) The form (\ref{Z}) ensures
that the metric function $e^\lambda$ behaves as
 \[ e^{\lambda} = 1 + {\it O}(r^2) \]
near $r=0$ for suitable choices of $n$. In fact this is a sufficient
condition for a static perfect fluid sphere to be regular at the
centre as pointed out by Maartens and Maharaj \cite{Manoj}. We
choose $n=-1$ in (\ref{Z}) to complete the integration. Additionally
we postulate the form
\begin{equation} \frac{E^2}{C} =
\frac{\alpha x}{(1+x)^2} \label{E}
 \end{equation}
  for $E$ which  depends on
the real valued parameter $\alpha$.  The form (\ref{E})  is
physically palatable since $E^2$ remains regular and positive
throughout the sphere if $\alpha
>0$. In addition the field intensity $E$ becomes zero at the stellar
centre and attains a maximum value of $E=\sqrt{\alpha C/2}$ when
$r=1/\sqrt{C}$. In what follows the Einstein-Maxwell  equations
(\ref{Durgc}) are solved for all nonnegative values of the parameter
$\alpha$.

With the choices (\ref{Z}) (taking $n=-1$)  and (\ref{E}),  equation
(\ref{Durgc-c}) reduces to
 \begin{equation}
  4(1+x)\ddot y -2\dot y + \left(1-\alpha\right)y = 0
\label{isotropy}
 \end{equation}
 It is convenient to categorise our solutions in
terms of different values of $\alpha$. We consider, in turn, the
following three cases:
 \[
\alpha=1, \,\,\, 0 \leq \alpha<1, \,\,\, \alpha>1
\]
which generate classes of solutions to the Einstein--Maxwell system
(\ref{Durgc}).

\subsection{The case $\alpha = 1$}

With the choice  $\alpha=1$, equation (\ref{isotropy}) assumes the
simpler form
\[
4(1+x)\ddot y -2\dot y = 0
\]
which is an ordinary differential equation of reducible order. It is
easily integrated to yield
\[
y=\frac{2}{3}c_1(1+x)^{3/2} + c_2
\]
where $c_1$ and $c_2$ are constants of integration to be determined
from physical considerations. The corresponding expressions for
$\rho$ and $p$ can then be established via (\ref{Durgc-a}) and
(\ref{Durgc-b}) respectively. The complete solution to the
Einstein--Maxwell field equations  (\ref{Durgc})
     is then given by
     \begin{subequations} \label{sol1}
\begin{eqnarray}
e^{2\lambda} &=& 1+Cr^2 \label{sol1-a} \\
 e^{2\nu} &=&
A^2\left[\frac{2}{3} c_1(1+Cr^2)^{3/2} +c_2\right]^2 \label{sol1-b}\\
\frac{\rho}{C} &=& \frac{Cr^2 + 6}{2(1+Cr^2)^2} \label{sol1-c} \\
\frac{p}{C}&=&\frac{c_1(10-Cr^2)(1+Cr^2)^{3/2}
-c_2(2+Cr^2)}{2(1+Cr^2)\left[ c_1(1+Cr^2)^{3/2} +3c_2 \right]}
\label{sol1-d} \\
E^2 &=& \frac{C^2r^2}{(1+Cr^2)^2} \label{sol1-e}\\  \sigma^2 &=&
\frac{C^2(3+Cr^2)}{(1+Cr^2)^5} \label{sol1-f}
\end{eqnarray}
\end{subequations}
Note that this charged solution does not have an uncharged analogue
as the electrostatic field intensity $E$ cannot vanish (except at
the centre). This effect essentially results from our condition that
$\alpha =1$. The line element for the solution (\ref{sol1}) is given
by
 \begin{equation}
ds^2 = -A^2\left[ \frac{2}{3} c_1(1+Cr^2)^{3/2} +c_2\right]^2dt^2 +
(1+Cr^2) dr^2 + r^2(d \theta^2 + \sin^2 \theta d \phi^2)
\label{metric1}
\end{equation}
In spite of the simplicity of this solution to the Einstein--Maxwell
equations, it does not seem to have been published before.

\subsection{The case $0\leq\alpha<1$}

With $0\leq \alpha <1$ equation (\ref{isotropy}) is more difficult
to solve. However it can be transformed to a standard Bessel
equation. It is convenient to introduce the substitution $ X=1+x$ in
(\ref{isotropy}) to yield
\begin{equation}
 4X\frac{d^2Y}{dX^2}-2\frac{dY}{dX}+(1-\alpha)Y =0 \label{isotropy2}
\end{equation}
 where $y(X)=Y$. We now introduce a new function $u(X)$ such that
$ Y(X)=u(X)X^m $ where $m$ is a real number. Then the ordinary
differential equation (\ref{isotropy2}) becomes
\begin{equation} 4X^2 \frac{d^2u}{dX^2}
+(8m-2)X\frac{du}{dX} +[4m^2-6m+(1-\alpha)X]u = 0 \label{isotropy3}
\end{equation}
 By introducing the further transform $  z = X^\beta $ where
$\beta$ is a real number, (\ref{isotropy3}) assumes the form
\[
4\beta^2X^{2\beta}\frac{d^2u}{dz^2}+\left[4\beta(\beta-1)X^\beta +
(8m-2)\beta X^\beta \right]\frac{du}{dz}  +
\left[4m^2-6m+(1-\alpha)X\right]u = 0 \] This equation may be
simplified by  the choice $\beta=\frac{1}{2}$ and $m=\frac{3}{4}$
which results in the differential equation
\begin{equation}
z^2\frac{d^2u}{dz^2}+z\frac{du}{dz}+\left[(1-\alpha)z^2
-\left(\frac{3}{2}\right)^2\right]u = 0 \label{isotropy4}
\end{equation}
 Now if we let
$(1-\alpha)^{1/2} z = w$ then (\ref{isotropy4}) becomes
\begin{equation}
w^2\frac{d^2u}{dw^2}+w\frac{du}{dw}+\left[w^2
-\left(\frac{3}{2}\right)^2\right]u = 0 \label{isotropy5}
\end{equation} which is the Bessel equation of order $\frac{3}{2}$.

The differential equation (\ref{isotropy5}) has linearly independent
solutions $J_{\frac{3}{2}}(w)$ and $J_{-\frac{3}{2}}(w)$ which are
Bessel functions. The general solution to (\ref{isotropy5}) can
therefore be written as
\begin{equation}
u=aJ_{\frac{3}{2}}(w)+bJ_{-\frac{3}{2}}(w)
 \label{u}
\end{equation} where  $a$ and $b$ are arbitrary constants. It is well known
that the Bessel functions of half-integer order can be written in
terms of the elementary trigonometric functions. In our case we
obtain the following explicit forms for $J_{\frac{3}{2}}$ and
$J_{-\frac{3}{2}}$ in terms of elementary trigonometric sine and
cosine functions:
\begin{eqnarray*} J_{\frac{3}{2}}(\sqrt{1-\alpha}
z)&=&\sqrt{\frac{2}{\sqrt{1-\alpha}\pi z}} \left[\frac{\sin
\sqrt{1-\alpha} z}{\sqrt{1-\alpha} z}-\cos \sqrt{1-\alpha} z \right]
 \nonumber \\
J_{-\frac{3}{2}}(\sqrt{1-\alpha}z)&=&-\sqrt{\frac{2}{\sqrt{1-\alpha}\pi
z}} \left[\frac{\cos \sqrt{1-\alpha} z}{\sqrt{1-\alpha} z}+\sin
\sqrt{1-\alpha} z \right]
 \end{eqnarray*}
 Then the general solution to the field
equation (\ref{Durgc-c}) may be written as
 \begin{eqnarray*} y(x) &=&
(1-\alpha)^{-3/4}\left[\left( c_1 +
c_2\sqrt{1+x} \right)\sin\sqrt{(1-\alpha)(1+x)} \right. \\
 & & \left. + \left(c_2-c_1\sqrt{1+x}
\right)\cos\sqrt{(1-\alpha)(1+x)}\right]
  \end{eqnarray*} where we have
introduced  the new constants $c_1 = a\sqrt{\frac{2}{\pi}}$ and $c_2
=-b\sqrt{\frac{2}{\pi}}$ for simplicity.

As $y(x)$ is now determined the forms for  $\rho$ and $p$ may then
be established via (\ref{Durgc-a}) and (\ref{Durgc-b}). The complete
solution of the Einstein--Maxwell system (\ref{Durgc}) for this
configuration, in terms of the radial coordinate $r$, is thus given
by the system
\begin{subequations} \label{sol2}
\begin{eqnarray}
e^\lambda &=& \sqrt{1+Cr^2}\label{sol2-a} \\
e^\nu &=&
\frac{A}{(1-\alpha)^{3/4}}\left[\left(c_1+c_2\sqrt{1+Cr^2}\right)
\sin f(r)  +\left(c_2-c_1\sqrt{1+Cr^2}\right)\cos f(r)\right]
\label{sol2-b}
 \\
\frac{\rho}{C} &=& \frac{(2-\alpha)Cr^2 + 6}{2(1+Cr^2)^2}
\label{sol2-c}\\  \frac{p}{C}&=& \left[ \left(\beta (2-\alpha)
(1+Cr^2)^{3/2}+(\alpha + 2 +4\tilde{\alpha})(1+Cr^2) +\beta(\alpha
+4\tilde{\alpha}) \sqrt{1+Cr^2}-\alpha \right)\right.  \nonumber\\
& & \left. +\left((\alpha-2)(1+Cr^2)^{3/2} +\beta (\alpha+2
+4\tilde{\alpha}) (1+Cr^2) \right. \right. \nonumber \\ & & \left.
\left. -(\alpha +4\tilde{\alpha})\sqrt{1+Cr^2}-\alpha
\beta\right)\tan f(r)\right] \times \nonumber \\ & &
\left[-2(1+Cr^2)^2\left( (\beta \sqrt{1+Cr^2}-1)-(\beta +
\sqrt{1+Cr^2})\tan f(r)\right)\right]^{-1} \label{sol2-d} \\
E^2 &=& \frac{\alpha C^2r^2}{(1+Cr^2)^2} \label{sol2-e} \\
\sigma^2 &=&\frac{\alpha C^2(3+Cr^2)^2}{(1+Cr^2)^5} \label{sol2-f}
\end{eqnarray}
\end{subequations}
 where we have set
  \[ \beta=\frac{c_1}{c_2}, \,\,\,
\tilde{\alpha}= \sqrt{1-\alpha}-1, \,\,\,  f(r)=
\sqrt{(1-\alpha)(1+Cr^2)}. \]
The line element for this solution has
the form
\begin{eqnarray} &&ds^2 =
-\frac{A^2}{(1-\alpha)^{3/2}}\left[\left(c_1+c_2\sqrt{1+Cr^2}\right)
\sin\sqrt{(1-\alpha)(1+Cr^2)} +\right. \nonumber \\
&& \left. \left(c_2-c_1\sqrt{1+Cr^2}\right)
\cos\sqrt{(1-\alpha)(1+Cr^2)}\right]^2 dt^2  +(1+Cr^2) dr^2 +
r^2(d\theta^2 +\sin^2\theta d\phi^2)\label{metric2}
\end{eqnarray} We believe that this model is
a new solution of the coupled Einstein--Maxwell system
(\ref{Durgc}). It is easy to verify that on setting $\alpha=0$ (i.e.
 $\tilde{\alpha}=0)$ we regain the Finch and Skea \cite{Finch} solution
for an uncharged sphere. From (\ref{sol2-e}) we have that $E = 0$ as
$\alpha =0$. The Finch and Skea solution is then given by
\begin{subequations} \label{Finch}
\begin{eqnarray}
e^\lambda&=&\sqrt{1+Cr^2} \label{Finch-a}\\  e^\nu&=& A\left[(c_1
+c_2\sqrt{1+Cr^2})\sin \sqrt{1+Cr^2} +(\:c_2-c_1
\sqrt{1+Cr^2}\:)\:\cos \sqrt{1+Cr^2}\: \right]
\label{Finch-b} \\
\rho&=&\frac{3+Cr^2}{(1+Cr^2)^2}\label{Finch-c}\\
p&=&-\frac{C}{1+Cr^2} \frac{ \left( \beta \sqrt{1+Cr^2}+1
\right)+\left( \beta- \sqrt{1+Cr^2} \right) \tan \sqrt{1+Cr^2}}{
\left( \beta \sqrt{1+Cr^2}-1\right) - \left( \beta+ \sqrt{1+Cr^2}
\right) \tan \sqrt{1+Cr^2}}\label{Finch-d}
 \end{eqnarray}
 \end{subequations}
  The uncharged solution
(\ref{Finch}) has been extensively studied by Finch and Skea and
shown to be regular in the interior of the star and matches smoothly
to the Schwarzschild exterior at the boundary. Furthermore, their
solution has been shown to be consistent with the neutron star
models proposed in the  the theory of Walecka \cite{Walecka}. Our
Einstein-Maxwell solution (\ref{sol2}) is a charged generalisation
of the physically reasonable model (\ref{Finch}) and does reduce to
it when $\alpha=0$.

\subsection{The case $\alpha>1$}

 We now briefly consider the case $\alpha>1$, so that
 $1-\alpha$ is negative, in the differential equation (\ref{isotropy}).
It is convenient to introduce the new constant $\psi$ so that
\[
1-\alpha = -\psi^2
\]
As the integration of (\ref{isotropy}) when $\alpha > 1$ is similar
to that in \S 3.2 we omit the details. The equivalent of equation
(\ref{isotropy4}) is
 \begin{equation} z^2\frac{d^2u}{dz^2} +z\frac{du}{dz}
- \left[\psi^2z^2+\left(\frac{3}{2}\right)^2\right]u = 0
\label{isotropy44}
 \end{equation}
  If we introduce the new independent variable
$\psi z = \tilde{w}$ then equation (\ref{isotropy44}) becomes
\begin{equation}
\tilde{w}^2\frac{d^2u}{d\tilde{w}^2} +\tilde{w}\frac{du}{d\tilde{w}}
- \left[\tilde{w}^2+\left(\frac{3}{2}\right)^2\right]u = 0
\label{591}
 \end{equation}
 We recognize (\ref{591}) as a special case of the modified Bessel
equation of fractional order. It has linearly independent solutions,
in terms of the modified Bessel functions $I_{\frac{3}{2}}$ and
$I_{-\frac{3}{2}}$, which may be expressed in terms of hyperbolic
functions. These are given by
 \begin{eqnarray*}
I_{\frac{3}{2}}(\psi z)&=&\sqrt{\frac{2}{\psi \pi z}}\left(
-\frac{\sinh (\psi z)}{\psi z}+\cosh (\psi z) \right)  \\
I_{-\frac{3}{2}}(\psi z)&=&\sqrt{\frac{2}{\psi\pi z}}\left(\sinh
(\psi z) -\frac{\cosh(\psi z)}{\psi z}\right)
 \end{eqnarray*}
  The general
solution to (\ref{591}) is thus given by
 \begin{equation}
u=aI_{\frac{3}{2}}(\psi z) + bI_{-\frac{3}{2}}(\psi z) \label{592}
 \end{equation}
  where $a$ and $b$ are constants. The equivalent of (\ref{592})
is (\ref{u}) for $0 \leq \alpha <1$. We can now  obtain the
gravitational potential $y(x)$ in the form
 \[
  y(x)= \left(c_1
\sqrt{1+x}-c_2 \psi\right) \sinh \left(\psi \sqrt{1+x}\right)   \\
    + \left( c_2 \sqrt{1+x}-c_1 \psi\right) \cosh \left(\psi
\sqrt{1+x}\right)
 \]
  which is the general solution of
(\ref{isotropy}) for $\alpha > 1$. Note that we have introduced the
new constants $c_1=b\sqrt{\frac{2}{\pi}}$ and
 $c_2=a\sqrt{\frac{2}{\pi}}$
 in $y(x)$.

The complete solution to the Einstein-Maxwell equations
(\ref{Durgc}) can now be determined. In terms of the radial
coordinate $r$ it is given by
 \begin{subequations} \label{sol3}
 \begin{eqnarray}
e^\lambda &=& \sqrt{1+Cr^2} \label{sol3-a}  \\
e^\nu &=& A\left[\left(c_1
\sqrt{1+Cr^2}-c_2 \psi\right)\sinh (\psi \sqrt{1+Cr^2})  \right. \nonumber \\
 & & \left.   + \left( c_2 \sqrt{1+Cr^2}-c_1 \psi\right)\cosh (\psi
\sqrt{1+Cr^2})\right] \label{sol3-b}\\
\rho&=&\frac{\left(1-\psi^2\right)C^2r^2+6C}{2(1+Cr^2)^2} \label{sol3-c}\\
\nonumber \\  p&=&\frac{(\psi^2-1)C^2r^2-2C}{2(1+Cr^2)^2}
+\frac{2}{1+Cr^2}\times \nonumber
\\  & &\frac{\left[ \psi^2 \left(  \tanh(\psi
\sqrt{1+Cr^2})+ \beta                          \right) \right]}{
\left[ \left( \beta \psi \sqrt{1+Cr^2}-1 \right) \tanh(\psi
\sqrt{1+Cr^2}) + \left(\psi \sqrt{1+Cr^2}-\beta \right)
              \right]}  \label{sol3-d}  \\
E^2&=&\frac{(\psi^2+1)C^2r^2}{(1+Cr^2)^2}  \label{sol3-e}   \\
\sigma^2 &=&\frac{(\psi^2+1) C^2(3+Cr^2)^2}{(1+Cr^2)^5}
\label{sol3-f}
 \end{eqnarray}
 \end{subequations}
  where we have put $\beta =c_1/c_2$ and $\alpha
=\psi^2+1$. The line element for this solution is given by
\begin{eqnarray}
ds^2 &=& A\left[\left(c_1 \sqrt{1+Cr^2}-c_2 \psi\right)\sinh(\psi
\sqrt{1+Cr^2})    + \left( c_2 \sqrt{1+Cr^2}-c_1 \psi\right)\cosh
\psi \sqrt{1+Cr^2}\right] dt^2   \nonumber \\
 && + (1+Cr^2)dr^2 +
r^2(d\theta^2 +\sin^2\theta d\phi^2)   \label{metric3}
 \end{eqnarray}
  It is interesting to
observe that this solution is similar to our charged analogue of the
Finch and Skea solution generated in \S 3.2. However in this case it
is not possible to eliminate the electric field $E$ (except at the
centre), in order to obtain an uncharged counterpart,
 since $\psi^2+1
>0$. This means that this solution models a sphere that is always
charged and hence cannot attain a neutral state.
 This feature is also shared by other charged solutions including that
 of Patel and Mehta \cite{Patel}.

 In summary the Einstein-Maxwell models represented by the interior
 metrics (\ref{metric1}), (\ref{metric2}) and (\ref{metric3}) are
 three new classes of exact solutions to the field equations which
 are all expressible in terms of elementary functions.

\section{Physical considerations: $0\leq\alpha<1$}

We now study the physical properties of the solution (\ref{sol2})
corresponding to  $0\leq\alpha<1$ which is the uncharged analogue of
the Finch and Skea neutron star model. Following Finch and Skea
\cite{Finch} we introduce the substitution
\[
v=\sqrt{1+Cr^2}
\]
in our solution (\ref{sol2}) and obtain the following forms for the
energy density $\rho$, the pressure $p$ and the electric field
intensity $E$:
\begin{subequations} \label{sol22}
\begin{eqnarray}
\frac{\rho}{C} &=& \frac{(2-\alpha)v^2 + \alpha +4}{2v^4} \label{sol22-a}\\
 \frac{p}{C}&=& \left[\left( (2-\alpha)\beta v^3 +(\alpha
-2+4\sqrt{1-\alpha})v^2 +(\alpha -4+4\sqrt{1-\alpha})\beta v
-\alpha\right)+ \right.\nonumber \\  & & \left. \left((\alpha -2)v^3
+ (\alpha -2+4\sqrt{1-\alpha})\beta v^2 - (\alpha
-4+4\sqrt{1-\alpha}) v-\alpha \beta\right) \tan
\sqrt{1-\alpha}v\right]  \nonumber \\ && \times \left[-2v^4
\left((\beta v -1)-(\beta
+v)\tan\sqrt{1-\alpha}v\right)\right]^{-1} \label{sol22-b} \\
\frac{E^2}{C}&=& \frac{\alpha(v^2-1)}{v^4} \label{sol22-c}
 \end{eqnarray}
 \end{subequations}
  in terms of the new variable $v$.

The gravitational behaviour of this model is difficult to analyse
because of the complexity of the expressions for $\rho$ and $p$. In
addition the solution is dependent on
 two parameters $\alpha$ and $\beta$; it is necessary to
select an appropriate value for one of the parameters and then
determine the second parameter  from physical considerations.
Specifying a value for $\alpha$ is not a simple matter. Some values
of $\alpha$ lead to unphysical behaviour (e.g.
 the value $\alpha = \frac{3}{4}$
results in the undesirable feature that $p_0=0$ for all choices of
$\beta$). Note that the subscript in $p_0$ denotes the centre $r=0$.
In the present discussion we select the value
\[ \alpha =
\frac{17}{81} \] for the physical analysis of the matter and
gravitational variables. This value of $\alpha$ chosen is consistent
with the interval of validity for $\alpha$ ($0 \leq \alpha < 1$) and
it can be demonstrated that this particular value of $\alpha$ does
not restrict the value of the parameter $\beta$ in general.
 The matter variables (\ref{sol22})
assume the following simpler forms for this choice of $\alpha$:
\begin{subequations} \label{581}
\begin{eqnarray}
\frac{\rho}{C}&=& \frac{145v^2 +341}{162v^4} \label{581a} \\
\frac{p}{C}&=& \frac{\left(145\beta v^3 +143v^2-19\beta v -17\right)
-\left(145v^3 -143 \beta v^2 -19v +17 \right) \tan\left(
\frac{8}{9}v\right)}{ 162v^4\left[(1-\beta v)+(\beta +v)\tan
\left(\frac{8}{9} v\right)\right]}
  \label{581b}    \\
\frac{E^2}{C} &=&\frac{17(v^2 -1)}{81v^4} \label{581c}
 \end{eqnarray}
 \end{subequations}
  It is
interesting to note that the Einstein-Maxwell solution (\ref{581})
satisfies a barotropic equation $p = p(\rho)$ which may be
determined explicitly. The pressure $p$ is given in terms of the
energy density by
 \begin{eqnarray}
\frac{p}{C}&=&\left( \left[145 \tilde{\rho}^{3/2} +143 \tilde{\rho}
-19 \tilde{\rho}^{1/2}
 -17\right]\right. \nonumber \\  &-& \left.\left[145
\tilde{\rho}^{3/2}
 -143
\tilde{\rho}-
 19
\tilde{\rho}^{1/2}
 +17 \right] \tan\left(
\frac{8}{9} \tilde{\rho}^{1/2}
 \right)\right) \nonumber \\  &\times&
\left( 162 \tilde{\rho}^2
 \left[\left(1-
\tilde{\rho}^{1/2}\right) +
  \left(1 +
\tilde{\rho}^{1/2}
 \right)\tan \left(\frac{8}{9}
\tilde{\rho}^{1/2} \right)\right]\right)^{-1}         \label{582}
 \end{eqnarray}
  where we have set
\[
\tilde{\rho} = \left( \frac{145C +\sqrt{21025C^2 +220968C\rho}}
{324\rho}\right)^{1/2}
\]
for convenience. Thus (\ref{582}), with $\alpha = \frac{17}{81}$,
represents a simple equation of state for a charged star.

 With  the  value of $\alpha= \frac{17}{81}$ the rate of change of
the energy density, pressure and the electric field intensity are
given by:
 \begin{subequations} \label{583}
 \begin{eqnarray}
\frac{d\rho}{Cdv}&=& \frac{-682 -145v^2}{81v^5} \label{583a}\\
\frac{dp}{Cdv}&=& \left( v\left[1-\beta v+(\beta +v)
\tan\left(\frac{8}{9}v\right)\right] \left[ 435\beta v^2 +286v
-19\beta \right. \right. \nonumber \\ & & \left. \left. -(435v^2
-286\beta v-19)\tan\left(\frac{8}{9}v\right)
 -\frac{8}{9}(145v^3 -143 \beta v^2
-19v +17 \beta) \sec^2\left(\frac{8}{9}\right) v)\right] \right.
\nonumber
\\  & &   \left.
 -
\left[145\beta v^3 +143v^2 -19\beta v-17
 -(145v^3 -143\beta v^2 -19v +17\beta)\tan
\left(\frac{8}{9}v\right)\right]\times  \right. \nonumber \\ & &
\left. \left[4-5\beta v +(4\beta +5v)\tan\left(\frac{8}{9}v\right)
+\frac{8}{9}v(\beta +v)\sec^2 \left.
\left(\frac{8}{9}v\right)\right]\right)  \right/  \nonumber \\ & &
\left(v^{5}\left[1-\beta v +(\beta +v)\tan\left(\frac{8}{9}v\right)
\right]^{2}\right)   \label{583b}   \\
\frac{1}{\sqrt{C}}\frac{dE}{dv} &=& \frac{\sqrt{\alpha}(2-v^2)}
{v^3\sqrt{v^2-1}} \label{583c}
 \end{eqnarray}
 \end{subequations}
 Using (\ref{583}) we obtain the
following expression for the speed of sound:
\begin{eqnarray}
\frac{dp}{d\rho}&=& \left( v\left[1-\beta v+(\beta +v)
\tan\left(\frac{8}{9}v\right)\right] \left[ (435\beta v^2 +286v
-19\beta) \right. \right. \nonumber \\  & & \left. \left. -(435v^2
-286\beta v-19)\tan\left(\frac{8}{9}v\right)
 -\frac{8}{9}(145v^3 -143 \beta v^2
-19v +17 \beta) \sec^2\left(\frac{8}{9}\right) v)\right] \right.
\nonumber
\\  & &   \left.
 -
\left[145\beta v^3 +143v^2 -19\beta v-17
 -(145v^3 -143\beta v^2 -19v +17\beta)\tan
\left(\frac{8}{9}v\right)\right]\times  \right. \nonumber \\ & &
\left. \left[4-5\beta v+(4\beta                \left.
+5v)\tan\left(\frac{8}{9}v\right) +\frac{8}{9}v(\beta +v)\sec^2
\left(\frac{8}{9}v\right)\right]\right) \right/ \nonumber \\ & &
\left((-290v^2 -1364)\left[(1-\beta v +(\beta
+v)\tan\left(\frac{8}{9}v\right) \right]^{2}\right)
\label{584}
 \end{eqnarray}
For a regular model we require that $\frac{d\rho}{dv}$,
$\frac{dp}{dv}$ and $\frac{dE}{dv}$ are well behaved in the interior
of the  charged star.

 To ensure that the star is regular at $r=0$ it is important to
examine the behaviour of our model at the centre of the star. The
quantities (\ref{581}), (\ref{583}) and (\ref{584}) have the
following forms at the centre of the star:
 \begin{subequations} \label{585}
 \begin{eqnarray}
\frac{\rho_0}{C}&=& 3 \label{585a}\\
\frac{p_0}{C}&=& \frac{2.232\beta -0.104}{\beta +9.620} \label{585b}\\
\left(\frac{d\rho}{Cdv}\right)_0&=& \frac{-827}{81}\label{585c}\\
\left(\frac{dp}{Cdv}\right)_0 &=& -\frac{363.79\beta^2 +1198.15\beta
+772.6}{(0.232\beta+2.232)^2} \label{585d}\\
\left(\frac{dp}{d\rho}\right)_0&=&\frac{0.22\beta^2 +0.723\beta
+0.905} {(0.232\beta +2.232)^2}
\label{585e}
 \end{eqnarray}
 \end{subequations}
  The quantity $E$
vanishes at $r=0$. In order to guarantee the positivity of the
central pressure we obtain the following restriction on $\beta$ from
(\ref{585b}):
\[
\beta \leq -9.62 \hspace {0.5cm} \mbox{or} \hspace{0.5cm} \beta \geq
0.104
\]
The adiabatic sound speed condition $0\leq \frac{dp}{d\rho}\leq 1$
yields the additional constraint:
\[
-4.06 \leq \beta \leq 5.9
\]
for the speed of sound not to exceed the speed of light. The
intersection of the above inequalities is given by
\[
0.104 \leq \beta \leq 5.9
\]
for a regular centre. It is clear that for this range of $\beta$ the
requirements $\rho_0 >0$, $\left(\frac{d\rho}{dv}\right)_0<0$  and
$\left(\frac{dp}{dv}\right)_0 <0$
 are trivially satisfied.

 Now we  are in a position to investigate the gravitational
behaviour of our model in the interior of the star. The behaviour of
the model is illustrated best in terms of graphs of the matter
variables and the gravitational potentials. These graphs have been
generated with the
 assistance of the software package Mathematica.
Based on the above limits obtained on $\beta$, we select the value
\[
\beta =1.
\]
The graphs of the various quantities are on the interval $1 \leq v
\leq 1.6$. Figures 1-3 represent the behaviour of the energy density
$\rho$, the pressure $p$ and the electric field intensity $E$,
respectively. In Figure 4 we have plotted $\frac{dp}{d\rho}$ on the
interval $1 \leq v \leq 1.6$. The metric functions $e^{2\nu}$ and
$e^{2\lambda}$ are plotted on the same interval in Figures 5 and 6,
respectively.

\begin{figure}[thb]
\vspace{1.8in} \includegraphics{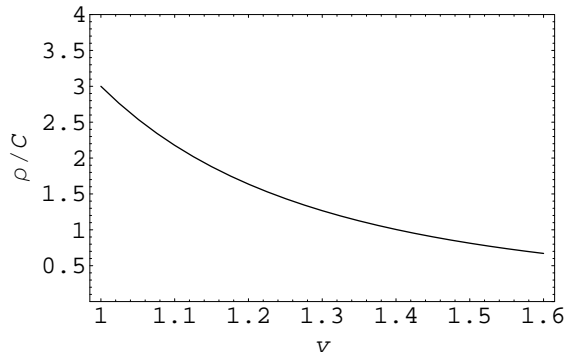} \caption{\label{Fig1}A plot of the
energy density $\rho/C$ versus $v=\sqrt{1+Cr^2}$.}
\end{figure}

\begin{figure}[thb]
\vspace{1.8in} \includegraphics{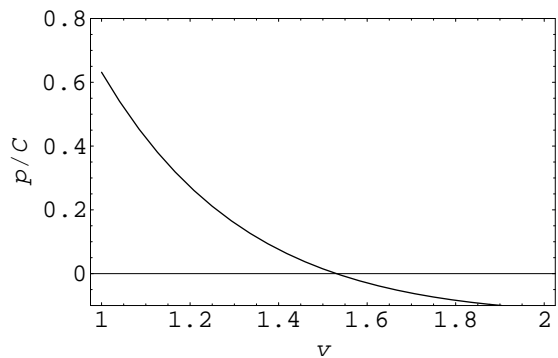} \caption{\label{Fig2} A plot of the
pressure $p/C$ versus $v=\sqrt{1+Cr^2}$.}
\end{figure}

\begin{figure}[thb]
\vspace{1.8in} \includegraphics{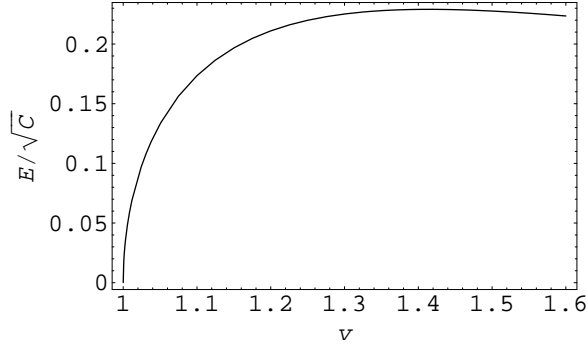} \caption{\label{Fig3} A plot of the
electric field intensity $E/ \sqrt{C}$ versus $v=\sqrt{1+Cr^2}.$ }
\end{figure}

\begin{figure}[thb]
\vspace{1.8in} \includegraphics{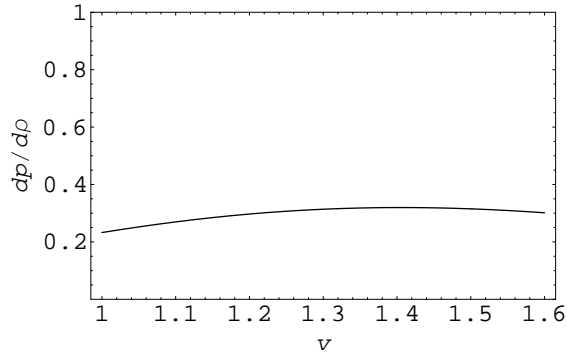} \caption{\label{Fig4} A plot of the
speed of sound $dp/d\rho$ versus $v = \sqrt{1+Cr^2}$.}
\end{figure}

\begin{figure}[thb]
\vspace{1.8in} \includegraphics{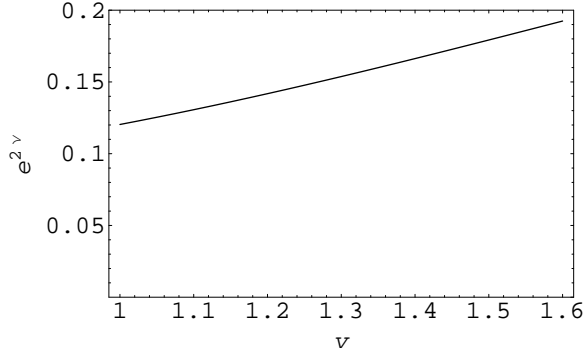} \caption{\label{Fig5} A plot of the
gravitational potential $e^{2\nu}$ versus $v = \sqrt{1+Cr^2}$.}
\end{figure}

\begin{figure}[thb]
\vspace{1.8in} \includegraphics{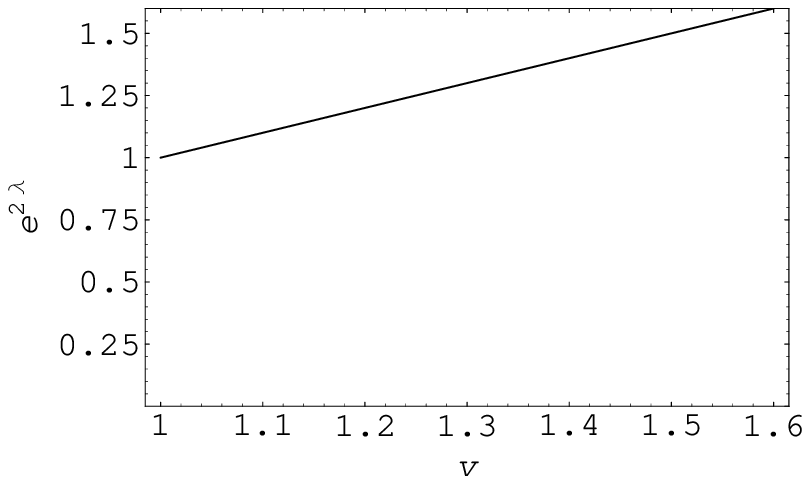} \caption{\label{Fig6} A plot of the
gravitational potential $e^{2\lambda}$ versus $v=\sqrt{1+Cr^2}$. }
\end{figure}

From the plots in Figures 1-2, we observe that the energy density
$\rho$ and  the pressure $p$
 are positive and monotonically decreasing functions
in the interior of the star. We observe from Figure 3 that the
electric field is positive and monotonically increasing on $1\leq v
< \sqrt{2}$, has a maximum value at $v = \sqrt{2}$, and then
decreases very slowly to the boundary. The potentials $e^{2\nu}$ and
$e^{2\lambda}$ are regular in the interior of the star as
illustrated in Figures 5 and 6 respectively. Thus the quantities
$\rho$, $p$, $E$, $e^{2\nu}$ and $e^{2\lambda}$ are continuous,
regular and well behaved throughout the interval $1 \leq v \leq
1.6$. From Figure 2 we note that the pressure vanishes at
approximately
 \begin{equation}
 v = 1.5252    \label{586}
  \end{equation}
   which
fixes the boundary of the charged star. The value (\ref{586}) is
consistent with the value of $v$ for which $p(v) =0$ where $p$ is
given by (\ref{585b}). Of course the boundary of the star will
change for other values of $\alpha$ and $\beta$. A pleasing feature
of this model is the behaviour of the speed of sound. It can be
observed from Figure 4 that the speed of sound is always less than
unity everywhere for $1 \leq v \leq 1.5252$.
 Therefore the causality
principle $0 \leq dp/d\rho \leq 1$ holds throughout the star and the
speed of sound is always less than the speed of light. Thus our
solution (\ref{sol2}), with $\alpha = \frac{17}{81}$ and $\beta =1$,
satisfies the requirements for a physically reasonable charged star.

 Utilising  the information obtained from the graphs we are now
able to examine the boundary conditions. In particular we have to
match the interior (\ref{metric2}) to the Reissner-Nordstrom
exterior (\ref{exterior}). With the help of $v = 1.5252 =
\sqrt{1+CR^2}$, the vanishing surface pressure condition $p(R)=0$,
yields the following approximate dependence of the constant $C$ on
the radius $R$ of the star:
 \begin{equation}
  C=\frac{1.33}{R^2} \label{586'}
  \end{equation}
   The continuity of the
electric field across the boundary condition $E(R) = Q/R^2$ gives
the  relationship of the total charge $Q$ of the star, as measured
by an observer at infinity,  with the radius $R$ for $\alpha =
17/81$:
 \[
Q = \frac{\sqrt{17}R^3}{9(1+R^2)}
\]
 Then we are in a position to find the  charge to radius
ratio
 \begin{equation}
  \frac{Q}{R} = 0.26  \label{587}
   \end{equation}
    On    matching of the
interior gravitational potentials $e^{2\nu}$ and $e^{2\lambda}$ from
(\ref{metric2}) with the exterior Reissner--Nordstrom solution
(\ref{exterior}) we obtain
 \begin{subequations}
\begin{eqnarray}
\frac{1}{1+CR^2} &=& 1-\frac{2M}{R} +\frac{Q^2}{R^2} \label{588a}\\
\frac{729}{512}c_1\left[(1+\sqrt{1+CR^2})\sin\left(\frac{8}{9}\sqrt{1+CR^2}
\right) \right. + && \nonumber \\ \left.
(1-\sqrt{1+CR^2})\cos\left(\frac{8}{9}\sqrt{1+CR^2}\right)\right]^2
&=& \frac{1}{1+CR^2} \label{588b}
 \end{eqnarray}
 \end{subequations}
 On substituting
(\ref{586'}) in (\ref{588b}) we find the following values for the
constants $c_1$ and $c_2$:
\[
c_1= 0.05444 = c_2
\]
while (\ref{588a}) gives another expression for $C$:
 \begin{equation}
C=\frac{2MR-Q^2}{R^2(R^2-2MR +Q^2)} \label{588'}
 \end{equation}
  in terms of
$Q$. Note that on setting $Q =0$ we obtain the same value of $C$ as
in the uncharged model of Finch and Skea \cite{Finch}. Therefore the
introduction of the electromagnetic field has the effect of
decreasing the value of $C$; consequently the radius $R$ of the
charged star is affected via the relationship (\ref{586'}).
Substituting for $C$ from (\ref{586'}) and for $Q$ from  (\ref{587})
in (\ref{588'}) we obtain the following mass-radius relationship:
 \begin{equation}
  \frac{M}{R} = 0.3  \label{587'}
  \end{equation}
   Note that   the Buchdahl
limit \[ M/R < 4/9 \] for stability is applicable for neutral stars.
We observe that this limit is not violated in our solution
(\ref{sol2}) in the presence of charge. In addition from (\ref{587})
and (\ref{587'}) we obtain the following mass-charge ratio
\[
\frac{M}{Q} = 1.154
\]
which satisfies the condition for equilibrium \[ M^2 > Q^2 \] of
charged stars with nonzero pressure in general as pointed out by
Cooperstock and de la Cruz \cite{Cooperstock}.

\section{Conclusion}
In this paper we obtained three categories of solutions for the
Einstein-Maxwell system (\ref{Durgc}) corresponding to the electric
field intensity (\ref{E}). In the first category  of solutions $E
\neq 0$ (except at the centre) and the solutions can be written in
terms of elementary functions. In the second category of solutions
it is possible for $E = 0$ throughout the interior spacetime and we
consequently regain the Finch and Skea solution \cite{Finch} which
is a good model for a neutron star. The charged analogue of the
Finch-Skea solution is given in terms of Bessel functions of
half-integer order. We demonstrated that this charged solution
satisfies an explicit barotropic equation of state. In the third
category of solutions $E \neq 0$ (except at the centre) and the
solutions can be written in terms of modified Bessel functions of
half-integer order.  A detailed analysis of the charged analogue of
the Finch-Skea solution was carried out. For specific parameter
values the energy density $\rho$, the pressure $p$, the electric
field intensity $E$, the speed of sound $dp/d\rho$, the
gravitational potential $e^{2\nu}$ and the metric function
$e^{2\lambda}$ were plotted. The profiles of the matter,
electromagnetic and gravitation variables suggest that these
quantities are well behaved so that the charged analogue of the
Finch-Skea solution describes a realistic charged stellar body. It
was shown that the speed of sound is less than the speed of light
and consequently causality is not violated. In particular we observe
that the mass-radius ratio is within the Buchdahl limit for
uncharged stars and the mass-charge ratio is consistent with the
requirements of Cooperstock and de la Cruz \cite{Cooperstock}. As
this model contains the physically acceptable Finch and Skea
solution and appears to be physically viable, a detailed analysis of
the stability should be carried out; this is the subject of ongoing
research.

\section*{Acknowledgements}
SH thanks the University of KwaZulu-Natal for a scholarship. SH and
SDM thank the National Research Foundation of South Africa for
financial support.

\end{document}